\documentclass[twocolumn]{emulateapj}

\newcommand{\Ha}{H$\alpha$}

\newcommand{\NII}{[{\sc N$\,$ii}]}

\newcommand{\OIww}{\hbox{[O\,{\sc i}]$\lambda\lambda $6300,6364}}
\newcommand{\Haw}{\hbox{H\,$\alpha\lambda $6563}}

\newcommand{\NIIwb}{\hbox{[N\,{\sc ii}]$\lambda $6583}}
\newcommand{\NIIww}{\hbox{[N\,{\sc ii}]$\lambda\lambda $6548,6583}}

\newcommand{\SIIww}{\hbox{[S\,{\sc ii}]$\lambda\lambda $6717,6731}}
\newcommand{\iraf}{{\sc Iraf}}
\newcommand{\calacs}{{\sc Calacs}}

\newcommand{\gmoss}{{\sc GMOS-S}}
\newcommand{\ifu}{{\sc IFU}}
\newcommand{\gifu}{{\sc GMOS-IFU}}

\def\kms{$\mbox{km s}^{-1}$}
\def\perpix{$\mbox{pix}^{-1}$}

\def\deg{^\circ}

\slugcomment{Accepted for publication in ApJ Letters}

\shorttitle{Streaming Motions Towards the SMBH in NGC~1097}
\shortauthors{Kambiz Fathi et al.}

\begin{document}

\title{Streaming Motions Towards the Supermassive Black Hole in NGC~1097}

\author{Kambiz Fathi}
\affil{Physics Department, Rochester Institute of Technology, 85
Lomb Memorial Dr., Rochester, New York 14623, USA}
\email{kambiz@cis.rit.edu}

\author{Thaisa Storchi-Bergmann \& Rogemar A. Riffel}
\affil{Instituto de F\`\i sica, UFRGS,
Av. Bento Goncalves 9500, 91501-970 Porto Alegre RS, Brazil}

\author{Claudia Winge}
\affil{Gemini Observatory, c/o AURA Inc., Casilla 603, La Serena, Chile}

\author{David J. Axon \& Andrew Robinson}
\affil{Physics Department, Rochester Institute of Technology,
85 Lomb Memorial Dr., Rochester, New York 14623, USA}

\author{Alessandro Capetti}
\affil{INAF - Osservatorio Astronomico di Torino,
Strada Osservatorio 20, 10025 Pino Torinese, Italy}

\author{Alessandro Marconi}
\affil{INAF - Osservatorio Astrofisico di Arcetri,
Largo Fermi 5, I-50125 Firenze, Italy}

\begin{abstract}
We have used GMOS-IFU and high resolution HST-ACS observations to
map, in unprecedented detail, the gas velocity field and structure
within the 0.7 kpc circumnuclear ring of the SBb LINER/Seyfert 1
galaxy NGC 1097. We find clear evidence of radial streaming motions
associated with spiral structures leading to the unresolved ($<3.5$
parsecs) nucleus, which we interpret as part of the fueling chain by
which gas is transported to the nuclear starburst and supermassive
black hole.
\end{abstract}


\section{Introduction}
Mechanisms of mass transfer in active galaxies from galactic scales
down to nuclear scales, have been the subject of many theoretical
and observational studies (e.g., Shlosman, Begelman, \& Frank 1990;
Emsellem, Goudfrooij, \& Ferruit 2003; Knapen 2005). Simulations
have shown that non-axisymmetric potentials efficiently promote gas
inflow towards the inner kpc-scale regions  (e.g., Engelmaier \&
Shlosman 2004). Further, recent observations have revealed that
structures such as small-scale disks or nuclear bars and associated
spiral arms are frequently observed in the inner regions of active
galaxies, suggesting they could be the means to transport gas from
the kpc scale down to within a few tens of parsecs of the active
nucleus (e.g., Knapen et al. 2000; Emsellem et al. 2001; Maciejewski
et al. 2002; Crenshaw, Kraemer, \& Gabel 2003; Fathi et al. 2005).

The nearby (14.5Mpc) LINER/Sey~1 galaxy NGC~1097 is an important
``laboratory'' in this context. It hosts a large-scale bar as well
as a secondary bar within its 0.7 kpc circumnuclear ring (e.g., Shaw
et al. 1993). Since the discovery of its double peaked \Ha\
emission-lines (Storchi-Bergmann, Baldwin, \& Wilson 1993), there
have been several studies to follow the ``fate'' of the gas
accumulated in the centre (e.g., Storchi-Bergmann et al. 2003).

Recently Prieto, Maciejewski, \& Reunanen (2005; henceforth PMR05)
have presented high resolution ground-based near-infrared images
obtained with the Very Large Telescope. They reported the discovery
of spiral structures within the inner 300 pc of the galaxy, which
can be traced to within $\approx$10 pc of the nucleus (the limit of
their spatial resolution). PMR05 argued that these spirals trace
cool dust, and could be channels by which cold gas and dust are
flowing to the nuclear supermassive black hole (hereafter SMBH).

Here we present two-dimensional maps of the gas kinematics and
structure within the inner 0.5$\times$1.0 kpc of NGC~1097, which
appear to trace the fueling path to the unresolved nucleus. Our
study is based on  high signal-to-noise ratio spectra obtained with
the Gemini Multi Object Spectrograph (\gmoss) Integral Field Unit
(\ifu), which reveal the inner gas kinematics in unprecedented
detail. We use these data in combination with high resolution images
obtained with the Advanced Camera for Surveys of the Hubble Space
Telescope (HST-ACS) to cross-correlate the gas kinematics with the
morphology of the nuclear spiral structure.

\section{Observations}
Our two-dimensional spectroscopic observations of NGC~1097 were
obtained with the \gifu\ on the Gemini South Telescope, on January
the 8th 2005, using the R400\_G5325 grating and r\_G0326 filter
(Gemini project GS-2004B-Q-25, PI: Storchi-Bergmann). These spectra
cover the wavelength range 5600-7000 \AA\ at a spectral resolution
of R$\approx$3500 (85 \kms\ FWHM). We observed three consecutive
fields of angular dimensions $5\arcsec\times7\arcsec$ each, covering
in total 15\arcsec\ along the major axis of the galaxy (position
angle PA=129$\deg$) and 7\arcsec\ along the minor axis, sampled by
hexagonal lenslets with projected diameters of 0\farcs2. Three
exposures of 600 seconds were obtained for each field, slightly
shifted along PA=40$\deg$ in order to correct for detector defects.
The spectra were reduced using standard tasks in \iraf. Products of
the reductions are three wavelength and flux-calibrated data cubes,
resampled to rectangular arrays of $50 \times 70$ spectra, each
corresponding to an angular coverage of $0\farcs1 \times 0\farcs1$.

The HST-ACS image for NGC~1097 was obtained (GO~9782; PI: Axon)
using the HRC mode (0\farcs025 \perpix\ and a field of view of
$26\arcsec\times 29\arcsec$), through the FR656N filter. The ACS
image was reduced and flux calibrated using the standard \calacs\
pipeline.

\section{Results}

\subsection{Spectroscopy}
\label{sec:spectroscopy} We used the two-dimensional spectroscopic
observations to map the kinematics of the ionized gas in the central
$0.5\times 1$ kpc. Inspection of the individual spectra reveals the
presence of the \OIww, \Haw, \NIIww, and \SIIww\ emission-lines
throughout most of the observed field. The velocity field was
derived from the central wavelengths of Gaussians fitted to the
narrow \Ha\ component and to the \NII\ emission-lines, as these
lines have good signal-to-noise ratio even at the borders of the
fields. Within the inner 0\farcs5, we introduced 3 additional
Gaussian components to fit and remove the underlying double-peaked
broad component of the \Ha\ line (see for example,
Fig.~\ref{fig:spectrum}). However, this does not significantly
affect the velocity field derived from the narrow lines. Since the
velocity fields of \Ha\ and \NIIwb\ are identical within the
uncertainties ($\approx 10$ \kms), we chose the latter to to
represent the gaseous velocity field of the galaxy because of its
higher signal-to-noise ratio.
\begin{figure}
\plotone{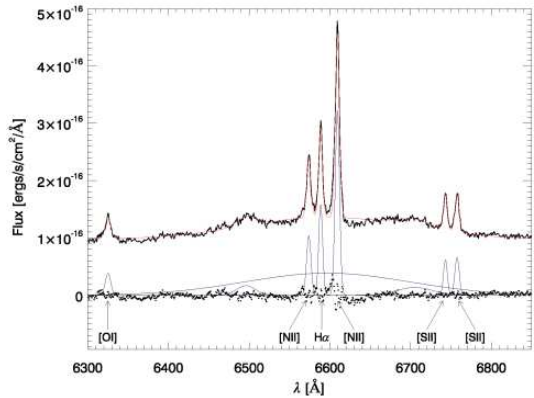} \caption{ Sample GMOS
spectrum of the nucleus illustrating the multi-Gaussian fit used to
determine the velocities of the narrow lines and remove the
underlying double-peaked broad H-alpha line. Each of the labelled
narrow lines is modelled by a single Gaussian profile. The broad
H-alpha line is represented by the 3 broader Gaussians. The
individual Gaussian components are plotted below the spectrum,
together with the residuals (dots) resulting from a simultaneous fit
to the spectrum of all the Gaussians plus a fixed continuum level.
The three ``unmarked'' Gaussians are a fit simply to remove nuisance
structure due to the double-peaked broad \Ha\ emission-line.}
\label{fig:spectrum}
\end{figure}

In the top panels of Fig.~\ref{fig:kinematics}, we present the
two-dimensional maps of the total flux  in the bandpass, the
integrated \NIIwb\ flux, and velocity dispersion ($\sigma$). The
nucleus is at position $(0,0)$, while a part of the circumnuclear
0.7 kpc star-forming ring is visible at the top of the panels. The
$\sigma$ map (corrected for the instrumental resolution) has
circumnuclear values around 40 \kms\ and decrease outwards to values
of $\approx$15 \kms, as one approaches the circumnuclear ring at
about 7\arcsec\ south-east from the nucleus.

The velocity field is shown in the bottom left panel of
Fig.~\ref{fig:kinematics}. It is clearly dominated by rotation with
distortions indicative of significant non-circular motions. Similar
distortions have previously been observed in the large-scale gas
kinematics of nearby disk galaxies (e.g. Visser 1980; Zurita et al.
2004; Emsellem et al. 2006), where they were found to trace the
galaxy spiral arms. In order to try to isolate these non-circular
motions we have fitted an exponential thin disk model (Freeman 1970)
to the velocity field using a method developed by Fathi (2004). In
the bottom middle panel of Fig.~\ref{fig:kinematics}, we illustrate
the best model, which has a deprojected maximum amplitude of
$\approx$200 \kms, a disk scale length of 3\farcs5, a kinematic PA
of $137\deg$ and disk inclination of $35\deg$. This inclination is
similar to that derived by Storchi-Bergmann et al. (2003) for the
nuclear accretion disk, but is smaller than the value of $46\deg$
found from the large-scale gas kinematics by Storchi-Bergmann et al.
(1996). The kinematic PA agrees with the value of $135\deg$ from
Storchi-Bergmann et al. (1996) and Emsellem et al. (2001). We have
tested the robustness of the model against extinction variations
across the field by repeating the fit several times, masking out the
regions most affected by the distortions. We found no significant
variations in the model parameters.
\begin{figure*}
\plotone{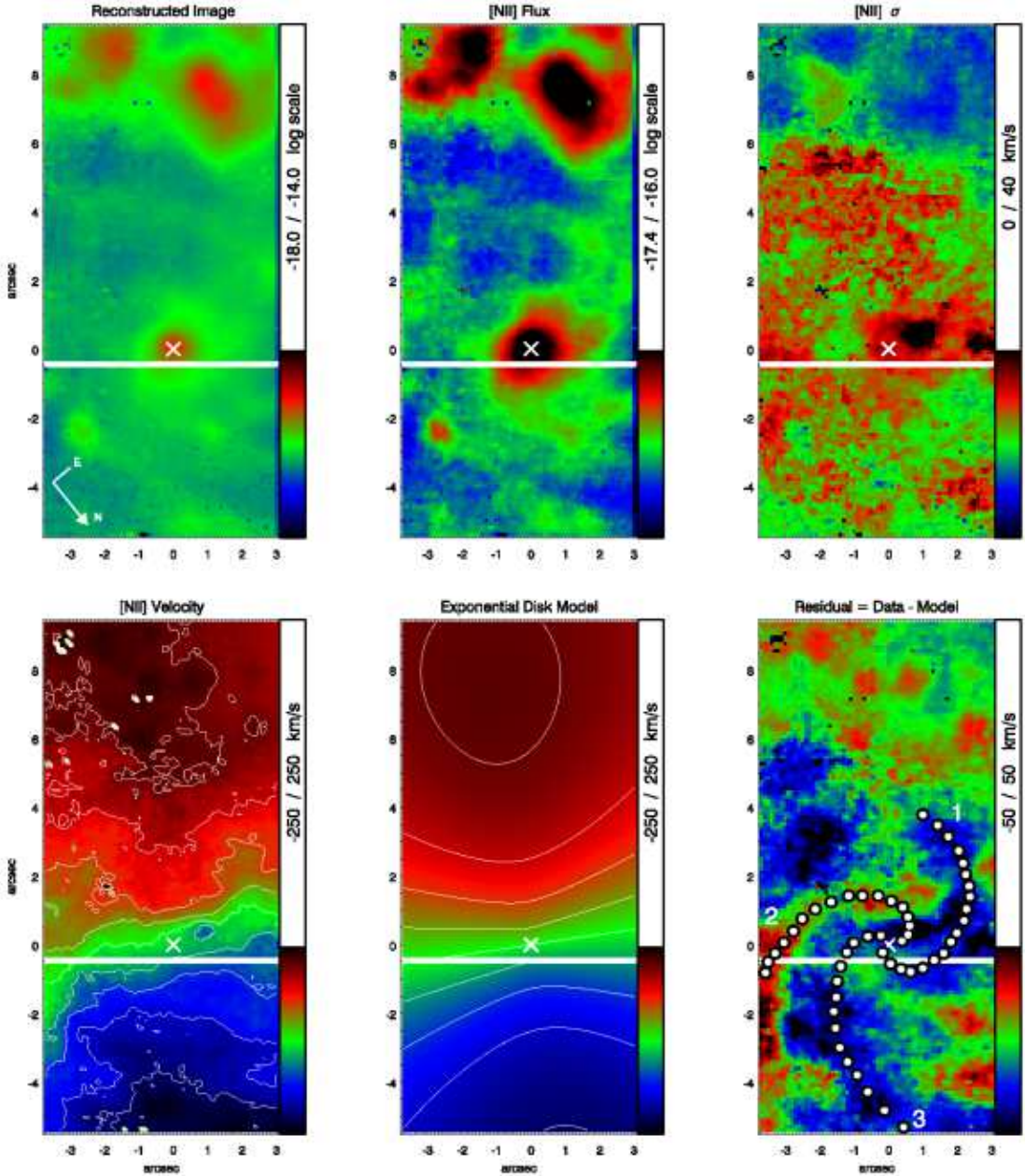} \caption{\gifu\ data results:
Reconstructed image formed by collapsing the 6300-6850 \AA\
wavelength range (top left); \NIIwb\ Flux distribution (top middle);
together with the kinematic maps derived from the \NIIwb\
emission-line, as well as the best fitting exponential disk velocity
field model and residuals. The spiral features are delineated by
white dots, with the numbers indicating the different arms as in
Fig.\ref{fig:acs}. In the bottom panels, red color indicates
redshift and blue color, blueshift. All panels share the same
orientation.} \label{fig:kinematics}
\end{figure*}

We constructed a residual map by removing the disk model from the
observed velocity field. The result is shown in the bottom right
panel of Fig.~\ref{fig:kinematics}, where it can be observed that
the residuals display amplitudes of $\approx$50 \kms. Moreover, the
velocity residuals are distributed in a spiral pattern, centered on
the nucleus. We identify three main arms, designated 1, 2, and 3,
which are outlined by white dots superposed on the residual map. The
residuals indicate redshifts along arm 2, but blueshifts along arms
1 and 3. As arms 1 and 2 are oriented approximately along the minor
axis, and the south-west is the near side of the galaxy
(Storchi-Bergmann et al. 1996), we conclude that the residuals
indicate radial inflow of gas along these circumnuclear arms. This
follows from the geometry of a configuration where the radial
motions occur along the spiral arms which are in the plane of the
disk. In this case, one expects the maximum line-of-sight velocities
along the minor axis. Furthermore, this implies that the spiraling
direction is clockwise. This simple setup does not fully explain the
observed blueshift along the inner parts of arm 3, where outflows or
individual gas clouds could complicate the observed velocities.

\subsection{Imaging}
\label{sec:imaging} We used two different techniques of image
analysis to investigate the structures present in the image. The
first is generating a ``structure map'', which is based on the
Richardson-Lucy image restoration technique (e.g., Pogge \& Martini
2002). The second method we have applied for enhancing the
morphological structures makes use of the \iraf\ {\small \sc
Stsdas/Ellipse} ellipse fitting routine to construct a model for the
galaxy. The structures are revealed in the residual image obtained
by subtracting the model from the original image (see
Fig.~\ref{fig:acs}).
\begin{figure*}
\plotone{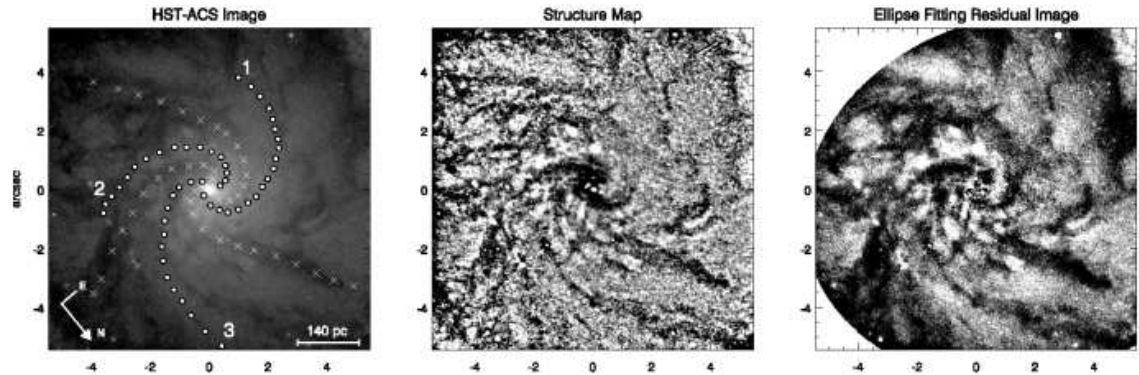} \caption{The inner $10\arcsec
\times 10\arcsec$ of the HST-ACS image of NGC~1097 (left), the
structure map (middle), and residual between the HST image and the
elliptical model (right). The angular sampling of 0\farcs025
\perpix\ corresponds to $\approx$1.8 pc \perpix\ at the galaxy. The
dots delineate the spiral arms found in our kinematic analysis,
labeled as in Fig.~\ref{fig:kinematics}, which follow clockwise
rotation (see Sec.~\ref{sec:spectroscopy}). The white crosses mark
the dark regions from Fig.~5 of PMR05.} \label{fig:acs}
\end{figure*}

The original ACS image shows several dark filaments, in the form of
spirals, which are enhanced in the processed images. A comparison
with the J-band image of PMR05 (their Fig. 5), shows good agreement
between the spiral structures seen in the optical and near-infrared,
and supports their interpretation that the dark filaments map dust
spirals which converge on the unresolved nucleus. The dark filaments
form a pattern similar to that seen in the velocity residuals.
However, a detailed comparison of the velocity residual map with the
ACS image shows that the spiral structure identified in the former
is rotated relative to that traced by the dark filaments in the
latter. To illustrate this, we have overplotted the loci of the
velocity residual spiral arms on the ACS image (Fig.~\ref{fig:acs}).
The radial streaming motions in the ionized gas occur {\em between}
the spiral dust lanes. The latter are indicated by white crosses in
Fig.~\ref{fig:acs}.

\section{Conclusions}
We have mapped the gas velocity field in the inner
7\arcsec$\times15$\arcsec of NGC~1097 with \gifu\ at an
unprecedented spatial resolution. These data have enabled us to
disentangle circular from non-circular motions in the ionized gas.
The dominant motion is disk rotation, upon which are superimposed
velocity residuals that delineate three spiral arms converging onto
the unresolved nucleus. We interpret the residual velocities - which
reach $\approx$20\% of the rotational velocity amplitude - as
evidence for radial inflow at $\approx$50 \kms\ from the region
interior to the circumnuclear 0.7 kpc ring, towards the nucleus.

A similar spiral structure delineated by dust lanes is seen in both
our optical HST-ACS image and the near-infrared images of PRM05. The
radial streaming motions in the ionized gas occur between the spiral
dust lanes. On larger scales, similar displacements between gas and
dust are common in galactic spiral arms (Tilanus \& Allen 1991).
This offset is attributed to loss of angular momentum of the gas as
it passes through the spiral shock (e.g., Combes 1996). Assuming
that the dust lanes trace the spiral shock, a similar mechanism may
operate in the inner kpc of NGC~1097. As gas initially on circular
orbits passes through the shock, some fraction of its kinetic energy
is thermalized, it loses angular momentum and falls toward the
center of the gravitational potential.

Taken at face value, the inferred streaming velocities would bring
the gas from scales of 10 pc (the spatial resolution of our
kinematic maps) down to the nuclear SMBH in approximately 200,000
yrs. Such a gas flow could also be responsible for triggering the
recently discovered starburst activity in the immediate vicinity of
the nucleus (Storchi-Bergmann et al. 2005).

NGC~1097 seems to have all the necessary features to allow gas
transfer from galactic scales down to the nucleus (e.g., Engelmaier
\& Shlosman 2004; Maciejewski 2004). It has a large-scale bar, which
is most likely the agent to bring gas down to the well known
circumnuclear ring. It has a secondary bar, which further funnels
gas toward the nuclear region, while here we have mapped streaming
velocities along spiral structures, down to at least 10 pc from the
nucleus. Further, the high resolution HST-ACS image suggests that
these structures continue down to $\approx$3.5 pc scale. It is the
first time that radial gas inflow has been mapped to such small
radii from the central supermassive black hole.

\acknowledgments Based on observations with the NASA/ESA Hubble
Space Telescope (HST-GO 09782.01) obtained at the Space Telescope
Science Institute, which is operated by the Association of
Universities for Research in Astronomy, Inc., under NASA contract
NAS5-26555. Also based on observations obtained at the Gemini
Observatory, operated by the Association of Univ. for Research in
Astronomy, Inc., under a cooperative agreement with the NSF on
behalf of the Gemini partnership: the NSF (USA), the PPARC (UK), the
National Research Council (Canada), CONICYT (Chile), the Australian
Research Council (Australia), CNPq (Brazil) and CONICET (Argentina).
T. Storchi-Bergmann acknowledges support from the Brazilian
institution CNPq.



\end{document}